# SEISMIC WAVE RECORDING BY 2S-SEISMOGRAPHS

By Ruhi Gurcan

**SUMMARY**

Researchers of seismic waves may construct a new seismographic recording adding one seismometer to each component of a conventional seismic station. The two identical conventional seismometers are set up in position of perpendicular and are connected in parallel feeding one recording device (digital or analog). This use of the seismometers (which they may be both horizontal or, one is vertical) is called "two seismometers seismograph" or simply "2S-S".

2S-seismograph performs new capabilities: 1.-it cause to a higher gain which is based on directly ground motion energy from the two orthogonal components of signals, 2.-it has a much smoother response curve than that of the single use of seismometer,3.-because of this smoothing, we are able to apply a higher level of static magnification which cause to widening the response at its both ends, therefore, 2S-System enable to work with a larger dynamic range frequency, 4.-it has a directional and motional filtering property which may be used in some cases advantageously, The contribution of "1", "2", "3" and "4" correspond to unique instrumental improvements for which seismography are ever needed.

Data which are obtained from the 2S-Ss have also more advantageous properties comparing with even that of ARRAY's: 5.-it is possible to record signals with their larger plane components all the time by a second 2S-S connected with the opposite ends, 6.-seismic wave types (P,S,R,L) can often be recorded separately on a separated 2S-seismogram since researchers usually deal with a known area of research, 7.-some implicit weak signals, which can not be readable as a phase on the conventional seismograms, become recorded newly and readably by the 2s-Ss.

In a non-directional 2S-recordings, 2S-sismograms contains the both seismic wave types being one predominant. However, 2S-seismograms which are obtained from the connection opposite ends include other wave types complementerly. Therefore, the two orthogonal 2S-seismogram contains more information and reading access than that of the conventional for all type waves, all the time and in any case due to “5”, “6”, “7” and “3”.

In addition to the above developments in seismographic instrumentation and the data obtained, as the construction of a 2S-sismic station is not as costly as ARRAY stations which uses hundreds of seismometer for the sake of a problem of better reading and detection , 2S-data may preferably gain applications used world-wide by an extensive users of seismic data.

## INTRODUCTION

Elastic waves which are classified based on the motion of individual particles with respect the directions of propagation of waves, we recognise as a phase on the seismograms. Knowledge about the structure of the Earth's interior and some quantities belong to earthquake itself are derived largely from the observational study of the seismic phases. Refinement of the obtained knowledge increases with the reading accuracy of the phases and along with the number of phases that are read.

In actual fact, a fairly large number of phases arrive at a seismic station from local and tele-seismic distances. In order to have an idea, I have drown figure 1 taking a two layers of structure with a buried source and a surface receiver at a distance of twice the layer thickness.

As shown in figure 1, we may receive more than twenty eight different phases even from such a simple two-layer crystal model. In the figure the phases starting with S-waves are not shown for the sake of simplicity. This simple model demonstrates that every discontinuity, elastic property of a layer and the quantities belong to earthquake may be evaluated several times for a better accuracy if the earth can be heard ideally.

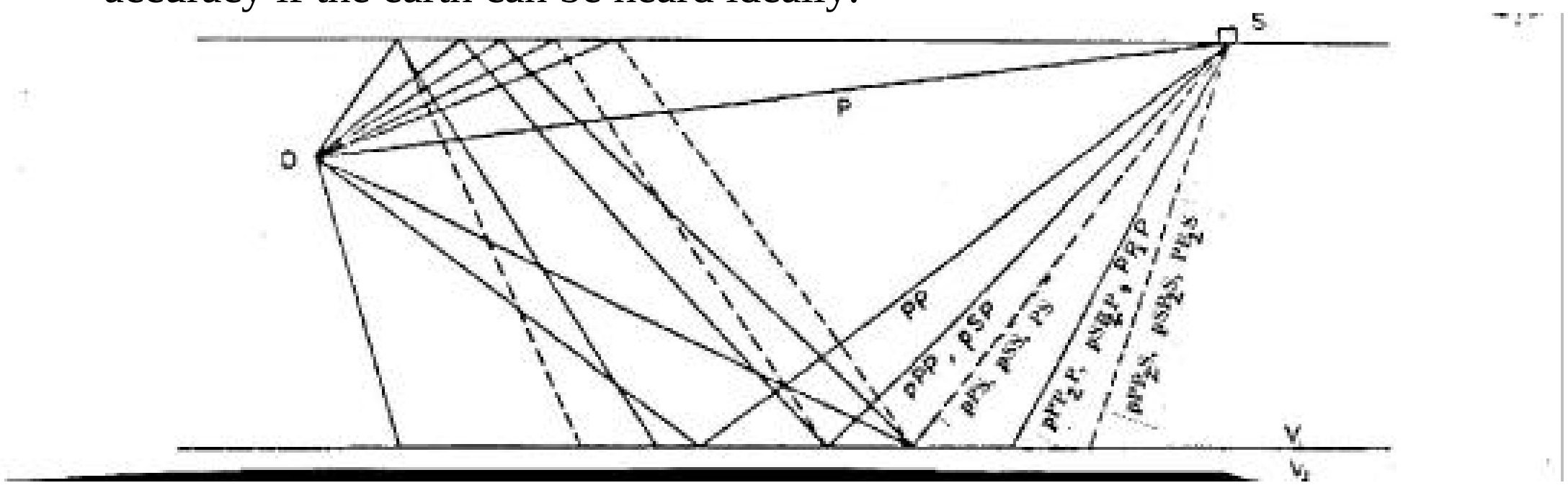

FIG.1. Phases involved in a layer over a half-space with a buried source.

On the other hand, due to the structure of the Earth, (with so many discontinuance in both crust and interior) a number of phases which are not readable today is increased because of ambient , seismic and instrumental noise. Also, insufficient seismographic detection capability and interfering phases of various kinds contribute to our inability in hearing the earth clearly. Therefore, simply increasing the sensitivity of an instrument is itself not sufficient to detect the signal’s existence and the kind of information gathered suggests that different recording configurations are desirable.

On the other hand, seismology being a observational science, researchers use various kind of approaches to discover recognizable phases from the conventional seismograms. In this respect, we may mention the works of White (1964), Shimshony and Smith (1964), Jurkevichs (1988), Samson (1981), Montalbetti and Kanasewich (1970), Vidale (1986) and others. In general , most of these techniques include procedures that adds or subtracts the perpendicular components of signals in some way mathematically for enhancement of signals read from the earlier conventional seismograms. However, all these work for a better recognition of phases are limited by the recording ability of the seismograms which are already recorded.

In this respect, 2S-seismograms constitute a ready and the most accurate data for these kind of works, shortening data acquisition since the subtraction or addition are provided by 2S-seismograms directly. One may favour that the combined signal components may be obtained from two channels by simply adding their digital traces. However, these data lack the accuracy of place, time and phase comparably as the two seismometers of the 2S-Seismograph are calibrated on purpose at the beginning for an equal output with the same equipment of loading, on the same pier before recording start. Additionally, it should be noted that, because of the 2S-Seismograph's instrumental abilities of the higher gain, the smoothing and the larger dynamic range, 2S-data will be never the same with the conventional in qualification.

The connection of seismometers in parallel actually has been previously treated by A.J. Serif (1959) in the literature. However, the positions of the two seismometers and their output are not taken into consideration for some possible detectional patterns of recording of seismic waves. For example, the seismometers had not been set up orthogonally and its linear output is not analysed for the purposes of detection of P-waves, S-waves, Rayleigh and Love waves which they all have distinct polarisation patterns for 2S-recording with the exception of some restrictive deviation in polarisation of S-waves as described by O. Nuttily (1962).

As the 2S-seismograms are the combination of two linear systems, the resultant output of the system must be analytically explicit, that is, linear. Therefore, I have also derived the transfer function of the 2S-S for basic elements earlier in1970 and published in1983.

For the derivation of the transfer function, first, the 2S-S is represented fully by an equivalent analog network where analogy is force-current and velocity-voltage is used, as proposed by Kolora and Russel (1966) secondly, network analysis and synthesis of the electrical engineering is used. in the calculation of output of 2S-system. What is important here is that using this analogy and analog representation I have also obtained the well known fourth and the fifth order equation of motion of the electromagnetic seismograph ( derived Savin and Carpenter, 1962) proving that the analogy chosen is appropriate. Upon these verifications " the sixth order equation of motion of an electromagnetic seismograph" is also derived by Garcon (1979).

Using the same analogy and representation for 2S-S, its transfer function is expressed by the ratio of polynomynal in term of Laplace parameter: S = iw ,and, the amplitude and the phase response are calculated and plotted for a frequency range for an impulse response Gurcan (1983). Importantly, from the plots seen that the magnification of the 2S-system which is changing based on the angle of azimuth and response curve of a seismometers are quite smoothed.

## THE EIGHT MAJOR CAPABILITIES OF THE 2S-SEISMOGRAPHS

The following eight major results may be given in an explanatory order for the 2S-seismograph: I- 2S-S perform the most direct and accurate combination of components of signals. II-2S-S provides larger plane components on the records rather than one directional component. III- 2S-S has directional and motional filtering property quadratically. IV- There exists the possibility of obtaining seismogram on which three dimensional P-or S-waves are recorded separately. V- 2S-S provides separated P-and S-wave type seismograms. VI-2S-S instrumentally provides much smoother response curves than that of conventional use of the seismometers. VII- 2S-S feeds one recording device with a high gain which is originated directly from the energy of detection.. VIII- An improvement in the dynamic range of seismographs are obtained by the smoothing and the applicability of high level static magnification..

Now, let us examine the way in which how these capabilities are achieved by the 2S-Seismographs.

I- **The direct and accurate combination**

All the signal polarisation combinations in the literature, are usually made from the data already recorded. This results in many inaccuracies. In this respect, 2S-Ss achieve higher level of accuracy because that: 1) Deliberately calibrated seismometers for an equality in the output of seismometers with some common loading at the beginning of its operation 2S-S, 2) The summation of the currents produced by the two seismometers with the common and equal instrumental constants is made directly and the most immediately for the seismometers for the detection before the recording is made.

II- **The larger plane components**

Since the recorded energy on the 2S-seismogram refer to the addition of two perpendicularly directional components of the signals and the plane coincides with the plane which actually is defined by the position of the two seismometers, the recording corresponds to a plane component of a signal. Note that as the addition of the components is not vectorial summation but the simple summation of the currents from the two seismometers the resultant plane component is larger gain-wise in strength than that of the vectorial combination. The relevant ratio may be written explicitly such as:

$$(\sin A + \cos A)/(\sin A.\sin A + \cos A.\cos A)^{1/2}$$

which corresponds to:

(the larger plane component) / (the vectorial plane component) .

This difference which is termed as “larger plane component” provides a great advantage just it is being larger .energy as much as forty percent compared to actual plane energy in representing the signals for recording against to noise and the friction. Thus, we must be aver of that the seismography gains an important tool in the resolution and recordability of weak signals on seismograms just because of this property of "larger plane component" .

## III-**Directional and motional filtering**

2S-seismographs may also be used as a directional and motional filter due to the ability of summing or extracting signal components, depending on the moving direction of senses of the seismometers' coil and the connection of their ends.

The currents produced by the two perpendicular seismometers exhibit two cases of mode in which the currents flow 1-) in the same direction or 2-) in the opposite direction. Thus, the modes are termed "in phase" or "out of phase" respectively.

These two cases of mode define two different quadrants of detection for seismic waves such as: the additional (+) quadrant of detection and the negational (-) quadrant of detection. This detectional patterns divide the surface of the earth into four quadrants, two of them is (+) and the other two is (-) quadrants for one type of seismic waves. The quadrants are defined by the direction of the booms of seismometers such as in the case of conventional where the axe of seismometer divides the Earth also into two hemispherically. Indeed, it is possible to cover all of the Earth's surface arrivals by both types of quadrants detectionally by setting a second pair of seismometers parallel to the first setting and connecting the ends of coils oppositely to the first one. Thus, no part of the Earth misses the advantages use of the additional (+) quadrants.

On the other hand, actually, each quadrant performs the filtration or enhancement depend on wave type with an angle of azimuth "A" which is the angle between the component of the particle motion of the seismic wave and the boom of the seismometers.

## IV_ **Recording signals three dimensionally**

There are two kinds of seismograms: one may simply be called "Horizontal" labelled "2S-H" where the two seismometers are both horizontal, and the other is called "Vertical" labelled "2S-V" where one component is horizontal and the second seismometer is vertical. The 2S-V works like a recording signals in three dimensional for P-waves when the "2S-H" is set up in a direction of focus. Actually, it is larger than the 3D because of the vertical component is added not vectoral but by the simple addition. This 2S-V works also as a matinal filter for the P-waves (or, filter the S-waves when it is set up perpendicular to the first one) The additional and negation quadrants replace each other depending on the connection of the seismometers' ends and also on the senses of the vertical component of signals. In order to receive signals of all wave types in the additional quadrants, a second 2S-V connected with the opposite ends is required.

## V_ **The separated P-wave S-wave seismograms**

It is well known that P-and S-waves, Rayleigh and Love waves all have distinct polarisation patterns and with a defined mutual relationships between their particle motion for homogeneously stratified earth.

Let us first consider the mutual relationship between S- and P-waves emanating from the same earthquake (with some exception for S-waves Nuttily , 1962). The S-wave particle motion, in general, is perpendicular to the P-wave's particle motion which coincides with the wave propagation path. In addition, the Rayleigh waves are elliptically polarised in the radial vertical plane while the Love waves are polarised rectilinearly in horizontal plane and orthogonal to the direction of the wave propagation. The Rayleigh type of surface waves may have horizontal components in the propagation direction, because of the elliptical movement of its particle motion, Nevertheless, they keep the orthogonally large extend with the motion of the Love waves.

Actually any deviation from these theoretical considerations must carry some knowledge from the region meaningfully and should be subjected to a investigation. In practice, particle motion polarisation of the seismic waves are rarely perfectly linear and orthogonal polorisationally because of the real world. But, this does not interrupt from an important result that seismic wave types fall detectionally into two different quadrants. That is, while one type of the wave components are added together in a (+) quadrant, the other type of wave components which are added in the next (-) quadrant. This results in the most direct and accurate technique of separating the P-and S-waves.

Note that the quadrants that are additional for one type of wave become negational quadrants for the other type of wave. This is true for both type of body and surface waves. Arrivals whose signal direction make an angle of about A=45 degrees with the boom of the seismometers will be eliminated almost completely from the seismograms when they are in the (-) quadrant or, oppositely signals will be recorded as much as doubly strengthen in the next quadrants compared to that of conventional seismograph's outputs. This doubling become true with the larger component combination. Here, another important point is that the two 2S-seismographs placed in parallel but connected at the opposite ends become complementary to each other. Thus, no information is missed at any time and even more information all the time is concerned with the 2S-S recordings. Therefore, we may have a seismogram on which waves coming from all quarters of the world are obtained with the enhancement or filtration for any type of wave.

VI_ **Smoothed response curves**

It is a fact that the response curves of a single seismometer seismographs (conventional uses) usually show a high notch at their resonance frequency on its response curves depending on its damping coefficient. However, with the use of a second seismometer on 2S-system, this high notch of the single seismometer which cause to a high dynamic magnification (for a narrow band of frequencies) is greatly smoothed out.

The existence of the second seismometer enlarges the circuit of the system and the current flow over both circuits belonging to the two seismometers. Whenever a heavy currents produced by one of the seismometers will be sent partly over the recording device while rest of it sent over the other seismometer causing a very important result such as smoothing in the response curve.

Therefore, another practical effect is that the second seismometer protects the recording device from the first seismometer's heavy current drivvings. Actually, this smoothing effect have been shown quantitatively by calculating the output from the transfer function of the 2S-system, for a unit impulse input, as mention in the part of introduction.

VII_ **The high gain**

Because of the signals are recorded directly by an energy of the ground vibration from the two perpendicular components, instead of one component, 2S-recording corresponds to a higher gain system compared to that of a single conventional use. Here, it is clear that this high gain is not obtained from an outsider connected feeder is connected for amplifying purposes but by an energy which is generated directly from the detection of the signals carrying knowledge about depth where the waves .pass through.

.

VIII_ **Broaden the dynamic range by the high level static magnification**

We have seen that the response curves of the 2S-seismograph are much more flatted than those of the single use of seismometers. Therefore, it becomes possible to apply a high level of static magnification over the input of a recording device. Obviously, this results in an elevation on the response curve whose useful part is widened at the ends for the both high and low frequencies. This makes the conventional seismometer work with a wider dynamic range than that of the single seismometer usages.

**DISCUSSIONS**

One of the important result of this connection of seismometers is to make weak signals readable, from which this result is obtained by the addition of the two orthogonal components of the signals. Secondly, as the signal components are not added vectorally but by their simple summation which makes the signals are represented energycally not only by their plane component but also larger in strength than that of signal's plane component itself in detection2. Thirdly, just because of this larger plane component, some signals which are not appeared at all on the conventional seismogram, may become readable as they able to overcome some of the noises and friction.

Actually, some more factors support the readability of the weak signals: such as the separating the signals according their wave types in directions and senses of arrivals, and by the applicability of high level statistical magnification through the smoothed the magnification over a wide range of frequency.

When we compare 2S-detection with that of ARRAY and telemetered stations, ARRAYS are being consist of a system of seismometers which are usually arranged in some regular geometric pattern over an area, it is difficult to accept that an array can act as a single station point of observation for providing a base to the particle motion of signals. Therefore, they will be of little help in the particle motion analysis of signals.

Arrays suppress a band of wavelengths to suppress any given noise by the phase tries serving to signal readings observationally. But , this result is achievable only when they are strong enough to be appeared against the noise on the seismogram. Secondly, in order to obtain high signal sensitivity by the combination of the outputs of many seismometers, an ARRAY requires an assumption that the pulse shape is at least approximately identical at the inputs of all the seismometers for a distance of kms. Actually, in many case some statistical communication theory must be adapted in order to produce a successful result from their data, Withcomb (1969), Ingate et all (1985).

In this respect, 2S-systems provide not only analytical plane polarised components of a signal at a point on the Earth for investigations but also discovering new phases from the weak signals which do not appear at all on the ARRAY’s seismograms due to noise and frictions.

The separation of the seismic waves by 2S-S provide clear shear-waves readings from the separated seismogram, which the waves typically contain three or four times the information carried by the P-wave train, Crampin (1985). However, although a clear and plane separated polarisation particle motion-wise, the azimuth studies is the weakest point in working with the 2S-seismographs. However, it is not difficult to obtain the E-W, N-S components and the angle of azimuth from the 2S-traces as their summation and negation are known.

The seismometers used in the 2S-seismograph may set up in a position ± 45 degrees to East for E-W and ±45 degrees to North for N-S components in order to get world-wide standard usage. However, this standardisation is not necessary for the most problem of seismology since many researchers deal with a known area of seismically active; since, many problem are related to the signals' existence and sensitivity rather than some statistical knowledge ; and, since, conventional directional component (EW,NS) recording of signals corresponds not to their decomposition provided by the Earth, but to a any division of one whole (particle) motion, why we don't measure signals with their larger plane components which provide the best readability and recognition for weak signals and a chance to discover new phases. The 2S-S that are directed toward an area for a solution of special problem, may also be useable for world-wide problem directly or applying some modification or reduction made by a computer program for 2S-data of standard settling.

In practice, the readability and using the short period seismometers as medium period seismometers will become extremely important since we use hundreds of seismometers in ARRAYS for a better readability.

Motional filtering and enhancement of waves potentially provides more information on readability of shear-wave splitting, especially, by the use of 2S-Vertical component. When one follow the senses of trace component on the 2S-V seismograms and some more details on particle motion may be obtained from the combination of seismometers connected with the opposite ends, The problem stated by Camden and Crampon (1991) "shear-wave arrivals may be contaminated by P-wave energy which will seriously distort information contained in the shear-wave splitting" may have chance to be investigated in more details.

Another related subject with the 2S-Ss, a question arises if there exist new design possibilities for electromagnetic 2S-seismographs because of the connection of two seismometers is always orthogonal. For example: (1)- having the two seismometers in one box, (2)- as feeding the recording device through two sources of seismometers, new construction may yield favourable reductions on the electro-dynamic motor constant normally, (3) it may also be useful in making the 2S-S more stable, and, (4) the existence of its own heavy electromagnetic damping resulting from the use of two seismometers may be used advantageously toward to manufacturing simplification by the proper choice of instrument parameters. Actually, they may all be useable together in making a compact form of 2S-S.

Perhaps, we should also add the possibility not to use some of the electronics which normally used with the conventional single seismometers in stations for filtration or amplification purposes. Therefore, some electronics do not necessarily need when 2S-system is used.

There is also some easiness in maintaining of 2S-seismographs because they work such as longer period of seismographs although seismometers connected are short period in fact. That is, the enlargement in the frequency response is effectively caused by 2S-S work like a longer period seismograph at the lover and higher side of its frequency of dynamic range for which actually the short period seismometers are used. Eventually, the operation and maintenance of the broader band 2S-S system will be as easy as that of these shorter period of seismometers' .

**2S-Recording Tests**

In order to make a precise comparison it would probably be necessary to place 2S-Ss operating alongside at a conventional station. An other important requirement would be the provision of the identical seismometers and recording devices (digital or analog) for the both types of seismographs. However, it may not be absolutely necessary identical the seismometers since this many instruments usually are not available at a seismic station for the 2S-tests.And, since, the small differences in instrumentation will not prevent us from showing the eight capabilities of 2S-system but hinder one from the quantitative studies. Finally, in ordinary, it is become difficult to find money to buy some extra identical seismometers without project for a seismic station.

However, as the arguments of 2S-Seismographs are so obvious, I would suggest a direct use of 2S-s for researchers who may advantageously use for solving their seismological problems. Actually, In order to 2S-Seismographs gain an applicability world–wide, a seismic station or an university, making a project, should dedicate to carry out 2S-recording experiments along with conventional records for one or two years and the observations should be published for all components.

Here, a limited amount of 2S-recording tests was made at The Technical University of Istanbul. Although a connection of two long period seismometers would have provided a more through picture of the functioning of the eight capabilities of 2S-S. The equipment available permitted for the 2S-experiments was the connection of two short period seismometers (Kinemetrix).

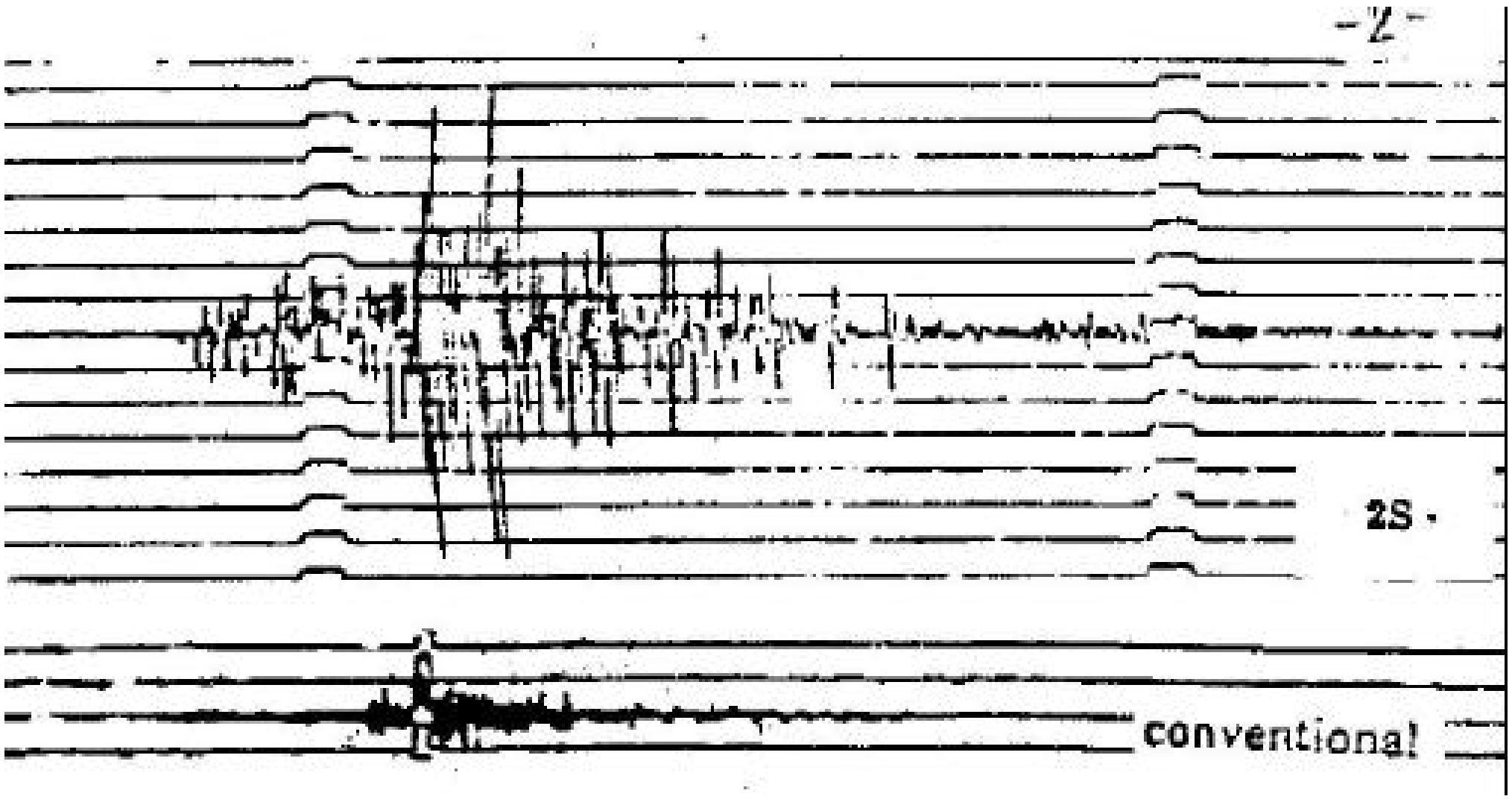

FIG.2 Records of the Sarayköy earthquake (37.97 N.28.77 E) March 25,1984, D =360 km. The upper portion of the record is from 2S-seismograph, the lower is from conventional seismograph (W.A.).

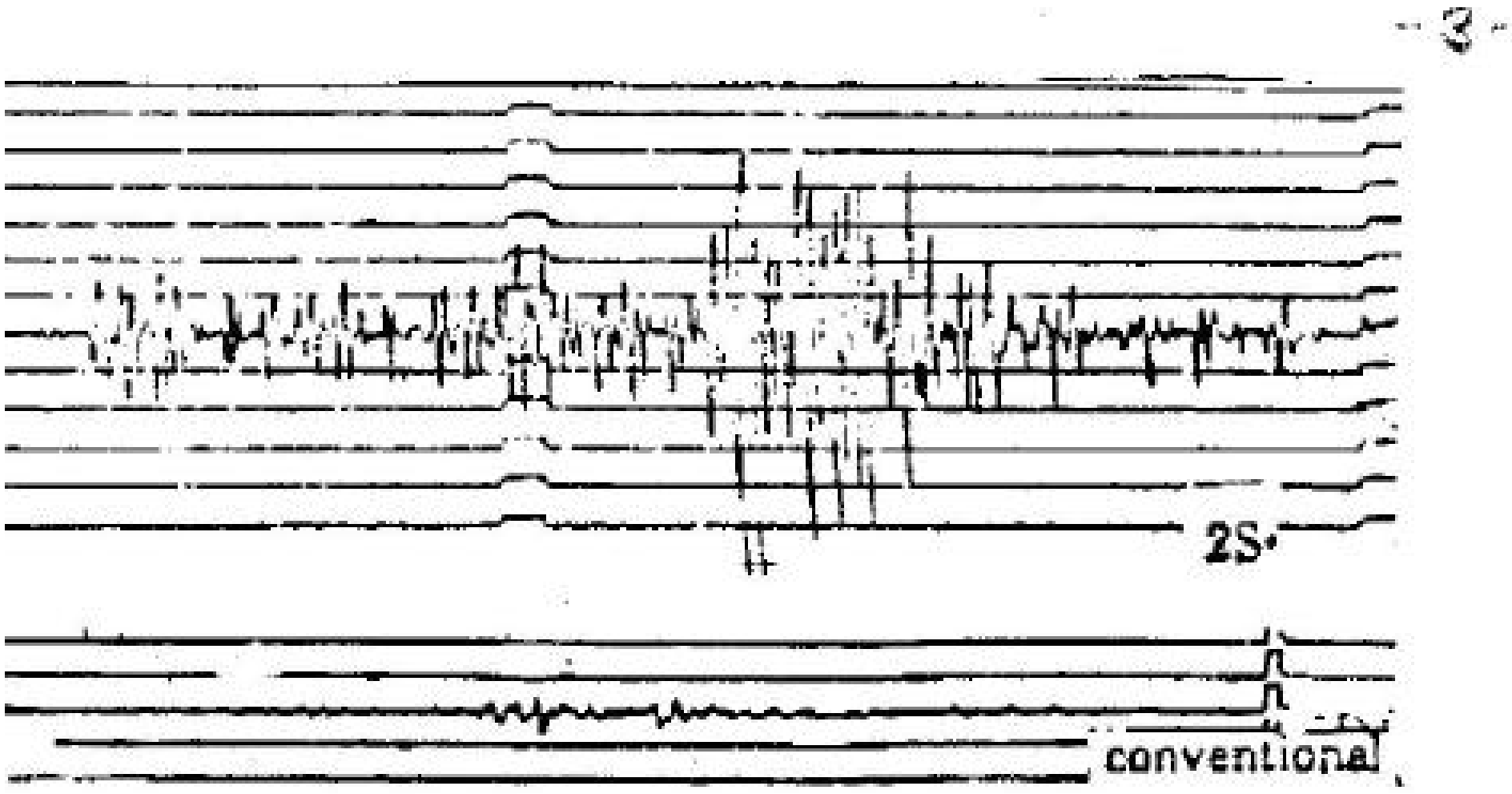

FIG.3 Records of the Adapazarý earthquake (40.68 N.30.45 E) March 26,1984. D =110 km. The upper record is from 2S-seismograph, the lower is from conventional seismograph.

The experiments lasted approximately one week. During this time, only two events could be taken into consideration. In order to make comparisons with conventional records of these two events, seismograms of Wood-Anderson were obtained from Kandilli, a seismic station a few km distant from the place where the 2S-seismograph experimental recordings were made. The portion of the conventional records which we want to compare are placed just the below the 2S-seismograms for easy comparison. See figures 2 and 3.

There has been no attempt for phase reading on either seismogram due to the lack of the large number of samples of data necessary for reliable identification at the begining.  Yet, it is possible to indicate some of the results from the 2S-seismograms.

The first seismogram records belong to Sarayköy earthquake.  The ratio of  the SH waves to the P-waves (SH/P) in amplitude measured from the 2S-seismogram are far greater than that of the E-W component of the conventional record.

It should be noted that, S-waves on the conventional record have larger components than P waves on the E-W seismogram because of the position of the hypocenter.  On the other hand, in the record of Adapazarý the same ratio of (SH/P) is of much less value when it is compared with the ratios taken from the conventional record even though, the E-W component of SH-waves is small due to the earthquake location. This reversal  result is obtained because the earthquake of Adapazarý was located approximately 80 degrees of azimuthal difference from the first earthquake.

The opposite ratio measurements show that two different quadrantal magnifications exist and that the arrivals which belong to the first earthquake approach "in the additional quadrant"  while similar wave arrivals (SH) of the second earthquake approach "in the negational quadrant"  of detection.

Again, I have to clarify that because of the instrumental possibilities, the records of these earthquakes do not provide a very good example to show the success of the 2S-seismographs. .For a simple  proving purposes yet, it should be sufficient to observe the summation and negation of the currents for the separation of the P-and S-waves as dominant characters on the 2S-seismograms. And, to show the large dynamic range,  is adequate a station test the both seismographs (2S-S and conventional) with identical seismometers on a shake-table for a large  range of frequency.

On the other hand, it should be born in mind that the experimental  proves will show only its practical side of 2S-seismographs, which it will always be possible to improve the outputs instrumentally especially in its early days.

## CONCLUSION

The connection of seismometers in the described form is not well documented in published literature. However, The output of the 2S-seismograph, being the summation of the output of two linear systems, is a linear output. This warrants analytical explanation which actually I studied in my early works (Garcon 1983). In this paper I tried to show some observational and practical usage of 2S-S. The 2S-seismograph with quite different abilities pertaining detection has  potential to attract the attention of researchers who may deal especially with a definite area of research in the earth science or even, in the exploration geophysics.

2S-seismograph combinations make contributions to major problems in seismography and in seismological data. Such that 2S-S provides: 1- the smoother response curves**,** 2- larger dynamic range, 3 -the higher detection gain for the weak signals, 4- the larger plane or 3D components 5-the greatest accuracy in combination of signals, 6-the capability of obtaining a seismograms which contain only P or S-waves' recordings, 7- discovering new phases, and 8- the recognising and recording ability for weakest phases on the seismogram comparable with arrays stations'.

By these capabilities of 2S-seismographs we have new potentialities not only some solution on the most important problems of seismography but an important results in reading and recognising weak signals which are not possible their records with the conventional recordings, including even, very expensive techniques such as arrays or seismographs linked by telemetry as they also lack the ability to record signals with the higher level dynamic range and the extra energy of larger plane components for recording against to noise.

On the other hand, as we have a new type of data, some modifications on algorithms which are being applied to the conventional seismograms for different purposes (such as for filtering and enhancing the signals over noise) may also be required. Fortunately, especially in the cases of polarisation (particle motion) and plane-component reductions, these modifications will shorten the calculation adding some accuracy to the results.

As a result of the above explanations become apparent that the 2S-Seismographs constitute new important capabilities for the science whose development is essentially based on observational success.

Acknowledgments

I thank Prof. Nezihi Canitez for permission to use the seismographs at Research Laboratories of Technical University of Istanbul. I also thank Mr. Uður Güllü and his colleagues for technical assistance in making 2S-tests.